\documentclass[12pt]{article}
\begin{document}
\begin{center}
\textbf{\Large Higher order corrections to the Grimus-Stockinger 
formula} \\[0.5cm]
S.~E.~Korenblit, D.~V.~Taychenachev \\ [0.5cm]
\end{center}
\begin{tabular}{l}
\textit{\indent Irkutsk State University, 664003, blwd Gagarin, 20, Irkutsk, Russia} \\
\textit{\indent e-mail: korenb@ic.isu.ru}
\end{tabular}
\begin{abstract}
\indent For the Grimus-Stockinger formula the same dimensionless parameter of asymptotic 
expansion is found by several ways of calculations. This parameter strongly 
depends on the width of wave packet.
\end{abstract}
\begin{tabular}{l}
\hbox{PACS: 14.60.Pq}\hfill \\
\end{tabular}
\newcommand{\q}{{\vec{\rm q}}}
\newcommand{\x}{{\vec{\rm x}}}
\newcommand{\n}{{\vec{\rm n}}}
\newcommand{\y}{{\vec{\rm y}}}
\newcommand{\vs}{{\vec{\rm v}}}
\newcommand{\p}{{\vec{\rm p}}}
\newcommand{\Pop}{{\mbox{\bf P}} }

\def\bldmth#1{%
\mathchoice
{{\hbox{\boldmath$\displaystyle#1$\unboldmath}}}%
{{\hbox{\boldmath$\textstyle#1$\unboldmath}}}%
{{\hbox{\boldmath$\scriptstyle#1$\unboldmath}}}%
{{\hbox{\boldmath$\scriptscriptstyle#1$\unboldmath}}}%
}
\def\vec#1{\bldmth{#1}}

\section{Introduction}

For the modern theory of neutrino oscillations \cite{Beuthe, NN} the main tool is 
Grimus-Stockinger theorem \cite{Grimus}, which gives the leading asymptotic behaviour 
with $|\vec{\rm R}|=R \rightarrow \infty$ for the integral:
\begin{equation}
\mathcal{J}(\vec{\rm R}) = \int \frac{d^3 q}{(2\pi)^3}\frac{e^{-i(\q\cdot\vec{\rm R})}
\Phi(\q)}{(\q^2 - \kappa^2 - i0)}\approx
\frac{e^{i{\kappa}R}}{4\pi R}\Phi\left(-\kappa\n\right)\left[1+ O(R^{-1/2})\right], 
\label{eq:integral}
\end{equation}
where: $\vec{\rm R} = R\n$, $\n^2=1$, and the function $\Phi(\q) \in C^3$ 
decreases at least like $1/\q^2$ together with its first and second derivatives. 
In order to understand the physical conditions necessary for this expansion the 
dimensionless parameters should be determined from the higher order corrections to 
this formula. Here this parameter is defined unambiguously by the use of various  
asymptotic expansions allowing to calculate the further corrections.

\section{Corrections for three-dimensional case.}

To obtain the higher corrections of order $R^{-n}$ we suppose that $\Phi(\q)$ and its 
first and second derivatives are represented by Fourier-transform as: 
\begin{equation}
\Phi(\q)=\!\int\! d^3{\rm x}\,e^{i(\q\cdot\x)}\,\varphi(\x), \quad 
\vec{\nabla}_q\Phi(\q)=i\!\int\! d^3{\rm x}\,e^{i(\q\cdot\x)}\,\x\varphi(\x), 
\mbox{ and so on.}
\label{2}
\end{equation}
Since $1/\q^2$ is also Fourier-image of $1/|\x|$, Eqs. (\ref{2}) are valid at least in the 
sense of distributions also for the functions defined above in \cite{Grimus}. 
By using the first equality of the following well known representations for spherical 
wave as a free Schr\"odinger 3- dimensional Green function with $\kappa=2\lambda$, $\q=2\p$: 
\begin{equation}
\frac{e^{i\kappa|\x|} }{4\pi|\x|}=\!
\int\!\frac{d^3{\rm q}}{(2\pi)^3}\frac{e^{\mp i(\q\cdot\x)}}{(\q^2-\kappa^2-i0)}=
i\!\int\!\frac{d^3{\rm p}}{4\pi^3}\,e^{\mp 2i(\p\cdot\x)}\!\int\limits^\infty_0\!dt\, 
e^{it(\lambda^2+i0-\p^2)},
\label{3}
\end{equation}
and by interchanging the order of integration for integral (\ref{eq:integral}) one 
finds: 
\begin{equation}
{\cal J}(\vec{\rm R})=\!
\int\! d^3{\rm x}\,\frac{e^{i\kappa|\vec{\rm R}-\x|}}{4\pi|\vec{\rm R}-\x|}\,\varphi(\x). 
\label{4}
\end{equation}
Substituting here the expansion, which in the exponential should always contain one 
additional order with respect to the ones in denominator: 
\[
|\vec{\rm R}-\x|=R\left[1-2\frac{(\n\cdot\x)}{R}+\frac{\x^2}{R^2}\right]^{1/2}=
R-(\n\cdot\x)+\frac{\x^2-(\n\cdot\x)^2}{2R}+\ldots ,
\]
we come to the corresponding expansion of integral (\ref{4}) up to $O(R^{-2})$:
\[
{\cal J}(\vec{\rm R})=\frac{e^{i\kappa R}}{4\pi R}
\int d^3{\rm x}\,e^{-i\kappa(\n\cdot\x)}\varphi(\x)
\left[1+\frac{(\n\cdot\x)}{R}+\frac{i\kappa}{2R}\left(\x^2-(\n\cdot\x)^2\!\right)+\ldots 
\right],
\]
that by making use of (\ref{2}) transcribes as:
\begin{eqnarray}
&&\!\!\!\!\!\!\!\!\!\!\!\!\!\!\!\!\!
{\cal J}(\vec{\rm R})=\frac{e^{i\kappa R}}{4\pi R}
\left[1-\frac i{R} (\n\cdot\vec{\nabla}_q)+
\frac{i\kappa}{2R}\left((\n\cdot\vec{\nabla}_q)^2-\vec{\nabla}^2_q\right)+
\ldots\right] \Phi(\q)\biggr|_{\q=-\kappa\n},
\label{5}  \\
&&\!\!\!\!\!\!\!\!\!\!\!\!\!\!\!\!\!
\mbox{with: }\;(\n\cdot\vec{\nabla}_q)\Phi(\q)\bigr|_{\q=-\kappa\n}=
-\partial_\kappa\Phi(-\kappa\n), \;\mbox{ and so on.}
\label{6} 
\end{eqnarray}
For any positive definite quadratic form of momentum $\q$: 
$\zeta=(\q\vec{\rm A}^{-1}\q)>0$,
\begin{eqnarray}
&&\!\!\!\!\!\!\!\!\!\!\!\!\!\!\!\!\!
\mbox{with: }\,
\Phi(\q)={\cal H}(\zeta), \;\;\,
\overline{\alpha}(\n)= \left(\n{\vec{\rm A}}^{-1}\n\right), \;\;\,
\overline{\alpha^2}(\n)= \left(\n{\vec{\rm A}}^{-2}\n\right),\,\mbox{ that is:}
\label{7} \\
&&\!\!\!\!\!\!\!\!\!\!\!\!\!\!\!\!\!
\left[1-\frac i{R} (\n\cdot\vec{\nabla}_q)+
\frac{i\kappa}{2R}\left((\n\cdot\vec{\nabla}_q)^2-\vec{\nabla}^2_q\right)\right]
\Phi(\q)\biggr|_{\q=-\kappa\n}\longmapsto
\nonumber \\
&&\!\!\!\!\!\!\!\!\!\!\!\!\!\!\!\!\!
\left[1+
\frac{i\kappa}{R}\left[3\overline{\alpha}(\n)-Tr\{{\vec{\rm A}}^{-1}\}\right]\!
\partial_\zeta-
\frac{i2\kappa^3}{R}\left(\overline{\alpha^2}(\n)-{\overline{\alpha}}^2(\n)\right)\!
\partial^2_{\zeta^2}\right]\!\!
{\cal H}(\zeta)\biggr|_{\zeta=\kappa^2\overline{\alpha}(\n)}.   
\nonumber
\end{eqnarray}
Then for Gaussian wave packet: ${\cal H}(\zeta) = e^{-\zeta/4}$, the expression (\ref{5}) 
reads: 
\begin{equation}
{\cal J}(\vec{\rm R})=\frac{e^{i\kappa R-\kappa^2{\overline{\alpha}}(\n)/4}}{4\pi R}
\left[1-\frac{i\kappa}{4R}\left[3\overline{\alpha}(\n)-Tr\{{\vec{\rm A}}^{-1}\}\right]-
\frac{i\kappa^3}{8R}\left(\overline{\alpha^2}(\n)-{\overline{\alpha}}^2(\n)\right)
\right].
\label{8}
\end{equation}
Here the square bracket evidently represents corrections only to the phase of  
the exponential. It may be directly obtained by the saddle-point method. 

To this end let's transcribe the integral (\ref{eq:integral}) for above Gaussian wave 
packet by using the second representation of Eq. (\ref{3}). Gaussian integration gives: 
\begin{eqnarray}
&&\!\!\!\!\!\!\!\!\!\!\!\!\!\!\!\!\!\!\!\!\!\!\!\!
{\cal J}(\vec{\rm R})=\frac{i}{4}\int\limits^\infty_0 dt 
\left[\frac{|{\vec{\rm K}}_t|}{\pi^3}\right]^{1/2} e^{iF(t)}, \quad 
\vec{\rm K}_t=\left({\vec{\rm A}}^{-1}+i t\vec{\rm I}\right)^{-1}, \quad 
\vec{\rm K}_0=\vec{\rm A},
\label{9} \\
&&\!\!\!\!\!\!\!\!\!\!\!\!\!\!\!\!\!\!\!\!\!\!\!\!
\mbox{where: }\,
iF(t)=it(\lambda^2+i0)-\left(\vec{\rm R}{\vec{\rm K}}_t\vec{\rm R}\right), 
\quad 
iF^{\prime\prime}(t)=
2\left(\vec{\rm R}\left\{{\vec{\rm K}}_t\right\}^3\vec{\rm R}\right), 
\label{10} \\
&&\!\!\!\!\!\!\!\!\!\!\!\!\!\!\!\!\!\!\!\!\!\!\!\!
iF^{\prime}(t)=i\left[\lambda^2+i0+
\left(\vec{\rm R}\left\{{\vec{\rm K}}_t\right\}^2\vec{\rm R}\right)\right]\mapsto 0, 
\quad t_0=R/\lambda +i\overline{\alpha}(\vec{\rm n})+\epsilon,
\label{11}
\end{eqnarray}
-- is the saddle point as asymptotical solution of Eq. (\ref{11}) for $R\to\infty$ up to 
the $\epsilon=O(\lambda/R)$. It is obtained by diagonalization 
$\vec{\rm A}=\vec{\rm O}^\top\,\overline{\vec{\rm A}}\vec{\rm O}$ onto the eigenvalues  
$\overline{\vec{\rm A}}={\rm diag}\{{\rm a}_j\}$, with $0<{\rm a}_j=1/\alpha_j<\infty$ 
and determinant
$|{\vec{\rm A}}|\equiv\det\{\overline{\vec{\rm A}}\}={\rm a}_1{\rm a}_2{\rm a}_3$, by 
using a suitable orthogonal rotation: $\vec{\varrho}=\vec{\rm O}\vec{\rm R}$, 
$\vec{\varrho}^2=\vec{\rm R}^2$, and due to Eq. (\ref{11}) defines $F(t_0)$ and 
$|{\vec{\rm K}}_{t_0}|$ up to $O(\epsilon^2)$. Along the path deformed according to  
$\overline{\alpha}(\n)>0$ we obtain:  
\begin{eqnarray}
&&\!\!\!\!\!\!\!\!\!\!\!\!\!\!\!\!\!\!\!\!\!\!\!\!
{\cal J}(\vec{\rm R})\approx \frac{i}{4}\,e^{i\pi/4}
\left[\frac{|{\vec{\rm K}}_{t_0}|}{\pi^3}\right]^{1/2}
\left[\frac{2\pi}{|iF^{\prime\prime}(t_0)| }\right]^{1/2}e^{iF(t_0)}=
\frac{e^{i\Theta(\vec{\rm R})}}{4\pi R}\,e^{-\lambda^2{\overline{\alpha}}(\n)},
\label{12} \\
&&\!\!\!\!\!\!\!\!\!\!\!\!\!\!\!\!\!\!\!\!\!\!\!\!
\Theta(\vec{\rm R})=2\lambda R-\,\frac{\lambda}{2R}
\left[3\overline{\alpha}(\n)-Tr\{{\vec{\rm A}}^{-1}\}\right]
-\,\frac{\lambda^3}{R}\left[\overline{\alpha^2}(\n)-{\overline{\alpha}}^2(\n)\right], 
\label{13} \\
&&\!\!\!\!\!\!\!\!\!\!\!\!\!\!\!\!\!\!\!\!\!\!\!\!
\mbox{that for: }\;\kappa=2\lambda,\quad Tr\{{\vec{\rm A}}^{-1}\}=\sum^3_{j=1}\alpha_j, 
\quad \sum^3_{j=1}\varrho^2_j\left[\overline{\alpha^n}(\n)-(\alpha_j)^n\right]=0,
\label{14}
\end{eqnarray}
exactly coincides with Eq. (\ref{8}) with the same precision. The corrections in 
(\ref{8}), (\ref{13}) evidently disappear for degenerate case:  
$\alpha_j=\alpha_1$, for $j=2,3$. 

For the neutrino oscillations problem: $\kappa=\sqrt{E^2_\kappa-m^2}\approx E_\kappa$, 
and for the Gaussian wave packet with coordinate width $\sigma_x$: 
$\vec{\rm A}\sim\sigma_x^{-2}$, so  
$Tr\{\vec{\rm A}^{-1}\}\sim\overline{\alpha}(\n)\sim\sigma_x^2$, whence, the true 
expansion parameters appear as combinations of two different dimensionless ones: 
$\kappa\sigma_x$ and $\sigma_x/R$, that define the application conditions of 
Grimus-Stockinger formula as:
\begin{equation}
(\kappa\sigma_x)\frac{\sigma_x}{R}\approx 
(E_\kappa \sigma_x) \frac{\sigma_x}{R}\ll 1,\;\mbox{ and: }\;
(\kappa\sigma_x)^3\,\frac{\sigma_x}{R}\approx 
(E_\kappa\sigma_x)^3\frac{\sigma_x}{R}\ll 1. 
\label{15}
\end{equation}
\section{The four-dimensional case.}
In fact the above integral (\ref{eq:integral}) is only three-dimensional part of the 
four-dimensional one defining macroscopic Feynman diagram \cite{NN} of the problem: 
\begin{eqnarray}
&&\!\!\!\!\!\!\!\!\!\!\!\!\!\!\!\!\!\!\!\!\!\!\!\!
J(R)
=\!\int\!\frac{d^4 q}{(2\pi)^2}\frac{e^{-i(qR)}\,\Phi(q)}{(q^2-m^2+i0)}=\!  
\int\!d^4r \frac {m^2}{i} h\left(i0-m^2(R+r)^2\right)\phi(r),
\label{eq:4int} \\
&&\!\!\!\!\!\!\!\!\!\!\!\!\!\!\!\!\!\!\!\!\!\!\!\!
\mbox{with: }\,R^\mu=(T,\vec{\rm R})\to\infty,\;\;
\sqrt{R^2}=\sqrt{R^\mu R_\mu}=\sqrt{T^2-\vec{\rm R}^2}\simeq T\frac{m}{E_\kappa}\leq T, 
\label{17} \\
&&\!\!\!\!\!\!\!\!\!\!\!\!\!\!\!\!\!\!\!\!\!\!\!\!
\mbox{where: }\,\frac{m^2}{i}h\left(m^2a^2\right)=
\!\int\!\frac{d^4 q}{(2\pi)^2}\frac{e^{-i(qx)}}{(q^2-m^2+i0)}\approx
\frac{m^2}{i}\sqrt{\frac{\pi}{2}}\frac{e^{-ma}}{(ma)^{3/2}}, 
\label{eq:4propagator}
\end{eqnarray}
for $a^2=i0-x^2=e^{i\pi}x^2$, is the causal propagator in coordinate space, and 
now the four-dimensional Fourier representation is assumed for $\Phi(q)$, 
which for relativistic Gaussian wave packet \cite{NN} reads as: 
\begin{eqnarray}
&&\!\!\!\!\!\!\!\!\!\!\!\!\!\!\!\!\!\!\!\!\!\!\!\!
\Phi(q)=\!\int\!d^4 r e^{-i(qr)}\phi(r),\;\;
\Phi(q)\mapsto e^{-(q\vec{A}^{-1}q)/4}, \;\; 
\phi(r)\mapsto \frac{|\vec{A}|^{1/2}}{\pi^2} e^{-(r\vec{A}r)}. 
\label{19}
\end{eqnarray}
Here again $\vec{A}^{-1}\sim \sigma_x^2$ in terms of Gaussian coordinate width
for any positively defined quadratic form of momentum $q$ in Minkowski space: 
$\zeta\!=(q\vec{A}^{-1}q)>0$. Then for $\sigma_x^2\to 0$ one has: $\vec{A}\to\infty$, 
$\Phi(q)\mapsto 1$, $\phi(r)\mapsto\delta_4(r)$, whence:  
$iJ(R)\mapsto m^2 h\left(i0-m^2R^2\right)$, that is reasonable from physical viewpoint.

Repeating now all the previous steps (\ref{4})--(\ref{8}) for the second expression 
(\ref{eq:4int}) of $J(R)$, with the so approximated propagator (\ref{eq:4propagator}), 
one obtains for arbitrary $\Phi(q)$ (\ref{19}) and    
$\eta^\mu=R^\mu/\sqrt{R^2}$, $l=i|l|$, $|l|=\sqrt{R^2}$, with
\begin{eqnarray}
&&\!\!\!\!\!\!\!\!\!\!\!\!\!\!\!\!\!\!\!\!\!\!\!\!
ma=ml\!\left[1+\frac{2i(\eta r)}{l}-\frac{r^2}{l^2}\right]^{1/2}\!\!\!
\approx m\!\left[l+i(\eta r)+\frac{(\eta r)^2-r^2}{2l}+\ldots\right]:
\label{20} \\
&&\!\!\!\!\!\!\!\!\!\!\!\!\!\!\!\!\!\!\!\!\!\!\!\!
J(R)\approx \frac{m^2}{i}\sqrt{\frac{\pi}{2}}\frac{e^{-ml} }{(ml)^{3/2}}
\left\{1+\frac{3(\eta\partial_q)}{2l}+\frac{m}{2l}\left[(\eta\partial_q)^2-
\partial_q^2\right]\right\}\Phi(q)\biggr|_{q=m\eta},
\label{eq:4diff}
\end{eqnarray}
that for the relativistic Gaussian wave packet from (\ref{19}), with 
$\overline{\alpha}({\eta})=\left(\eta{\vec{A}}^{-1}{\eta}\right)$, 
$\overline{\alpha^2}({\eta})=\left({\eta}{\vec{A}}^{-2}{\eta}\right)$, 
analogously gives:
\begin{eqnarray}
&&\!\!\!\!\!\!\!\!\!\!\!\!\!\!\!\!\!\!\!\!\!\!\!\!
J(R)\approx\frac{m^2}{i}\sqrt{\frac{\pi}{2}}\, \frac{e^{-ml}\,
e^{-m^2\overline{\alpha}(\eta)/4} }{(ml)^{3/2}}
\label{eq:4asymptotic} \\
&&\!\!\!\!\!\!\!\!\!\!\!\!\!\!\!\!\!\!\!\!\!\!\!\!
\cdot \left\{1-\frac{m}{4l}
\left[4\overline{\alpha}(\eta)-Tr\{\vec{A}^{-1}\}\right] -\frac{m^3}{8l}
\left[\overline{\alpha^2}(\eta)-\left(\overline{\alpha}(\eta)\right)^2\right]\right\}.
\nonumber
\end{eqnarray}
Such, on the first sight, rough calculations are exactly confirmed again by saddle point 
method. Indeed, by exponentiating like (\ref{3}) the denominator of the first expression 
(\ref{eq:4int}) with $m=2{\rm m}$ and representation dependent 
$\vec{g}\mapsto g_{\mu\nu}$ or $\delta^\mu_{\;\nu}$, by means of Gaussian integration 
clarified in Appendix one has instead of (\ref{9})--(\ref{11}): 
\begin{eqnarray}
&&\!\!\!\!\!\!\!\!\!\!\!\!\!\!\!\!\!\!\!\!\!\!\!\!
J(R)=\frac 1{i}\!\int\limits^\infty_0\!dt|\vec{K}(t)|^{1/2}
\exp\left\{-i{\cal F}(t)\right\},\quad 
\vec{K}(t)=\left[\vec{A}^{-1}-it\vec{g}\right]^{-1},
\label{23} \\ 
&&\!\!\!\!\!\!\!\!\!\!\!\!\!\!\!\!\!\!\!\!\!\!\!\!
-i{\cal F}(t)=-it\left({\rm m}^2-i0\right)-\left(R\vec{K}(t)R\right), \quad 
-i{\cal F}^{\,\prime\prime}(t)=2\left(R\vec{K}^3(t)R\right), 
\label{24} \\ 
&&\!\!\!\!\!\!\!\!\!\!\!\!\!\!\!\!\!\!\!\!\!\!\!\!
-i{\cal F}^{\,\prime}(t)=-i\left[{\rm m}^2-i0+\left(R\vec{K}^2(t)R\right)\right]\mapsto 0,
\quad t_0=|l|/{\rm m}-i\overline{\alpha}({\eta})+\epsilon,
\label{25} 
\end{eqnarray}
$t_0$ is again the saddle point as asymptotical solution of Eq. (\ref{25}) for 
$|l|\to\infty$ up to the $\epsilon=O({\rm m}/|l|)$. It is obtained now by diagonalization 
in Minkowski space as $\vec{A}=\vec{\xi}^{(j)}(\overline{\vec{A}})_{jn}\vec{\xi}^{(n)}$  
onto the eigenvalues $(\overline{\vec{A}})_{jn}= g_{jn} a_{(j)}$ ordered \cite{Synge} as  
$0<g_{jj}a_{(j)}=g_{jj}/\alpha_j<\infty$, with determinant
$|\vec{A}|=|A_{\mu\nu}|=|(\overline{\vec{A}})_{jn}|$, by using a suitable Lorentz 
transformation as $\varrho^j=\xi^{(j)}_\mu R^\mu$ with 
$\varrho_j=g_{jj}\varrho^j=g_{jk}\varrho^k$, $\varrho^2=R^2$, and due to Eq. (\ref{25}) 
defines again ${\cal F}(t_0)$ and $|{\vec{K}}(t_0)|$ up to $O(\epsilon^2)$.  
Along the path deformed according to $\overline{\alpha}(\eta)>0$, instead of 
(\ref{12}) one finds:
\begin{eqnarray}
&&\!\!\!\!\!\!\!\!\!\!\!\!\!\!\!\!\!\!\!\!\!\!\!\!
J(R)\approx \frac{1}{i}|\vec{K}(t_0)|^{1/2}
\left[\frac{e^{-i\pi/2}\,2\pi}{|-i{\cal F}^{\prime\prime}(t_0)|}\right]^{1/2}\!\!\!
e^{-i{\cal F}(t_0)}=
\frac{\sqrt{{\rm m}\pi}}{i}\frac{e^{-iB(R)}
e^{-{\rm m}^2\overline{\alpha}(\eta)} }{l^{3/2}},  
\label{26} \\ 
&&\!\!\!\!\!\!\!\!\!\!\!\!\!\!\!\!\!\!\!\!\!\!\!\!
-iB(R)=-2{\rm m}l-\frac{{\rm m}}{2l}
\left[4\overline{\alpha}(\eta)-Tr\{\vec{A}^{-1}\}\right]-\frac{{\rm m}^3}{l}
\left[\overline{\alpha^2}(\eta)-\left(\overline{\alpha}(\eta)\right)^2\right], 
\label{27} \\ 
&&\!\!\!\!\!\!\!\!\!\!\!\!\!\!\!\!\!\!\!\!\!\!\!\!
\mbox{that for: }\,
Tr\{\vec{A}^{-1}\}
=(\vec{A}^{-1})^\lambda_{\;\,\lambda}=\sum^3_{j=0}\alpha_j,\;\;\,
|A_{\mu\nu}|=\prod\limits^3_{j=0}g_{jj}a_{(j)}>0,
\label{28} \\
&&\!\!\!\!\!\!\!\!\!\!\!\!\!\!\!\!\!\!\!\!\!\!\!\!
\overline{\alpha^n}(\eta)=\!\sum^3_{j=0}(\alpha_j)^n g_{jj}\frac{\varrho^2_j}{\varrho^2}=
\left(\eta\vec{A}^{-n}\eta\right),\quad g_{\mu\nu}\,,g_{jk}={\rm diag}\{1,-1,-1,-1\}, 
\label{29}
\end{eqnarray}
exactly coincides with Eq. (\ref{eq:4asymptotic}) with the same precision. 

The true parameters of expansion appear again as the following products of the two 
dimensionless parameters, that are now $m\sigma_x$ and $\sigma_x/|l|$:
\begin{equation}
(m\sigma_x)\frac{\sigma_x}{|l|}\ll 1,\; \mbox{ and: }\; 
(m\sigma_x)^3\,\frac{\sigma_x}{|l|}\ll 1,
\label{eq:4condition} 
\end{equation}
and they have the same order for $(m\sigma_x)\leq 1$. 

It is easy to see that both the conditions for 3- and 4-dimensional 
cases are practically the same. Indeed, Eq. (\ref{17}) implies that 
\begin{equation}
\vs = \frac{\vec{\kappa}}{E_\kappa} = \frac{\vec{\rm R}}{T},\;\mbox{ whence: }\;
R^2=|l|^2= \frac{\vec{\rm R}^2}{\kappa^2}m^2=\frac{T^2}{E^2_\kappa}m^2.
\label{31}
\end{equation}
Since for ultrarelativistic neutrino $T\approx|\vec{\rm R}|$ and 
$E_\kappa \approx\kappa=|{\vec{\kappa}}|$, the 
conditions (\ref{eq:4condition}) may be rewritten as: 
\begin{eqnarray}
&&\!\!\!\!\!\!\!\!\!\!\!\!\!\!\!\!\!\!\!\!\!\!\!\!
(\kappa\sigma_x)\frac{\sigma_x}{|\vec{\rm R}|}=
(E_\kappa\sigma_x)\frac{\sigma_x}{T}\approx 
(E_\kappa\sigma_x)\frac{\sigma_x}{|\vec{\rm R}|}\ll 1,
\label{32} \\
&&\!\!\!\!\!\!\!\!\!\!\!\!\!\!\!\!\!\!\!\!\!\!\!\!
\mbox{and: }\;
(m\sigma_x)^2(\kappa\sigma_x)\frac{\sigma_x}{|\vec{\rm R}|}\approx
(m\sigma_x)^2(E_\kappa\sigma_x)\frac{\sigma_x}{|\vec{\rm R}|}\ll 1.
\label{33}
\end{eqnarray}
Thus, for $(m\sigma_x)\leq 1$ these both conditions are the same as the first one in 
3 - dimensional case (\ref{15}). Moreover the same dimensionless parameter (\ref{32}) 
defines in fact the asymptotical solutions of both the saddle points equations (\ref{11}) 
and (\ref{25}). Note that exact values of the first's and second's 
square brackets in (\ref{13}) and/or (\ref{27}) respectively may be different, and their 
determination in terms of $\sigma_x$ for 4 - dimensional case \cite{NN} 
(\ref{27})-(\ref{29}) is different from 3 - dimensional case (\ref{13}), (\ref{14}).

The authors thank V. Naumov, D. Naumov, E. Akhmedov, S. Lovtsov, and N. Iljin for 
useful discussions. 

\section{Appendix.}

In order to strictly calculate a standard Gaussian integral over the Minkowski space  
\cite{NN}:
\begin{equation}
\int d^4y\,\exp\left\{-(y\vec{A}y)+2(By)\right\},
\label{34}
\end{equation}
where the quadratic form $(y\vec{A}y) = y_\mu A^{\mu\nu}y_\nu$ is symmetric and positive definite, the following solution of the eigenvalue problem may be used: 
\begin{equation}
\vec{A}\vec{\xi}^{(n)} = a_{(n)}\vec{\xi}^{(n)},\;\;(\vec{\xi}^{(n)})^2=g^{nn},\;\;
A_{\mu\nu}=\xi^{(l)}_\mu(\overline{\vec{A}})_{ln}\xi^{(n)}_\nu, \;\; 
(\overline{\vec{A}})_{ln} = g_{ln} a_{(n)}.
\label{35}
\end{equation}
In spite of ambiguity (\cite{LL} \S 94) of diagonalization procedure for symmetric 
tensor in Minkowski space, the positive definiteness of $\vec{A}$ leaves the used type 
of its diagonalization only, leading to eigenvectors 
$\vec{\xi}^{(n)}$, $n=0\div 3$ (\ref{35}), whose components ${\xi}^{(n)}_\nu$ define 
Lorentz transformation diagonalizing the form. Then with the substitutions 
$Y^n=\xi^{(n)}_\nu y^\nu$ transforming $(y\vec{A}y)=(Y\overline{\vec{A}}Y)$, 
$b^m=B^\mu\xi^{(m)}_\mu$, $(By)=(bY)=b^mg_{mn}Y^n$, the integration (\ref{34})  
factorizes to:
\begin{eqnarray}
&&\!\!\!\!\!\!\!\!\!\!\!\!\!\!\!\!\!\!\!\!\!\!\!\!
\int\!\!d^4Y\!\exp\!\left\{\!-(Y\overline{\vec{A}}Y)+2(bY)\right\}\!\equiv\! 
\prod\limits^3_{n=0}\!\left\{\int\limits^\infty_{-\infty}\!\!dY^n\,
e^{-(Y^n)^2g_{nn}a_{(n)}+2b^ng_{nn}Y^n}\right\}\!= 
\nonumber \\
&&\!\!\!\!\!\!\!\!\!\!\!\!\!\!\!\!\!\!\!\!\!\!\!\!
=\sqrt{\frac{\pi^4}{|(\overline{\vec{A}})_{ln}|}}\,
\exp\left[\left(b\left(\overline{\vec{A}}\right)^{-1}b\right)\right]=
\sqrt{\frac{\pi^4}{|A_{\mu\nu}|}}\,\exp\left[\left(B\vec{A}^{-1}B\right)\right],\;
\label{36}
\end{eqnarray}
where $A_{\mu\nu}\left(\vec{A}^{-1}\right)^{\nu\lambda}\!=\delta_\mu{}^\lambda$, and 
$|A_{\mu\nu}|=|(\overline{\vec{A}})_{ln}|$ is defined by (\ref{28}). 

Nevertheless it is instructive obtain the same result without reference to 
diagonalization by using the direct integration over space and time variables separately. 
Since for n-dimensional Minkowski space with signature metric 
$g_{\mu\nu}={\rm diag}\{1,-1,-1,\ldots,-1\}$ in any given orthogonal basis the symmetric 
tensor $\vec{A}$ is represented by the block matrix $A_{\mu\nu}=A_{ij}$ for 
$i,j,\mu,\nu=0\div n-1$, with the rightmost bottom block 
${\cal A}_{ij}={\cal A}^{ij}={\cal A}_{ji}$, for $i,j=1\div n-1$:
\begin{eqnarray}
A_{\mu\nu}=A_{\nu\mu}=\!\left(\!\!\begin{array}{cc}
A_{00} & A_{0j}        \\
A_{i0} & {\cal A}_{ij}  
\end{array}\!\! \right),
\label{37}
\end{eqnarray}
the integral (\ref{34}) with $d^4y\mapsto d^n y$ may be rewritten as: 
\begin{eqnarray}
\!\int\limits^\infty_{-\infty}\!\!dy^0
e^{-y^0A_{00}y^0+2(B_0y^0)}\!\int\!d^{n-1}{\rm y}
\exp\left[-{\rm y}^l{\cal A}_{lk}{\rm y}^k-2\left(y^0A_{0k}-B_k\right){\rm y}^k\right].
\nonumber 
\end{eqnarray}
The both integrals are over Euclidian space now, so they are evaluated to
\begin{equation}
\sqrt{\frac{\pi^n}{\alpha|{\cal A}_{ij}|}}\exp\left\{\!
\left(\!B_l\left(\vec{{\cal A}}^{-1}\right)^{lk}\!\!B_k\!\right)+
\frac{1}{\alpha}
\left(\!B_0-B_l\left(\vec{{\cal A}}^{-1}\right)^{lk}\!\!A_{k0}\!\right)^2\right\},
\label{38}
\end{equation}
if $\alpha\equiv A_{00}-A_{0l}\left(\vec{{\cal A}}^{-1}\right)^{lk}\!\!A_{k0}>0$.
The expression is simplified by Laplace expansion of the determinant $|A_{\mu\nu}|$, 
where $M\left({}^{i_1}_{j_1}{\;}^{i_2}_{j_2}{\;}^{...}_{...}\right)$ means the minor 
of the matrix $A_{\mu\nu}$, whose rows $i_1, i_2, ...$ and columns $j_1, j_2, ...$ are 
deleted:
\begin{eqnarray}
&&\!\!\!\!\!\!\!\!\!\!\!\!\!\!\!\!\!\!\!\!\!\!\!\!
|A_{\mu\nu}|\equiv \sum^{n-1}_{k=0}(-1)^{k}A_{0k}M\left({}^0_k\right) =
A_{00}M\left({}^0_0\right)-\sum^{n-1}_{k=1}(-1)^{k+1}A_{0k}M\left({}^0_k\right)=
\nonumber \\
&&\!\!\!\!\!\!\!\!\!\!\!\!\!\!\!\!\!\!\!\!\!\!\!\!
= A_{00}M\left({}^0_0\right)-\sum^{n-1}_{k=1}(-1)^{k+1}A_{0k}\sum^{n-1}_{i=1}
A_{i0}(-1)^{i+1}M\left({}^0_0{\;}^i_k\right)=
\nonumber \\
&&\!\!\!\!\!\!\!\!\!\!\!\!\!\!\!\!\!\!\!\!\!\!\!\!
= |{\cal A}_{lj}|\left[A_{00}-A_{0k}\left(\vec{\cal A}^{-1}\right)^{ki}A_{i0}\right] = 
\alpha|{\cal A}_{lj}|. 
\label{39}
\end{eqnarray}
Thus $\alpha>0$ due to positivity condition of the form (\ref{37}) implying that 
$|A_{\mu\nu}|, |{\cal A}_{lj}|>0$. Furthermore, if for any symmetric block matrix 
$\vec{A}$:
\begin{eqnarray}
&&\!\!\!\!\!\!\!\!\!\!\!\!\!\!\!\!\!\!\!\!\!\!\!\!
\vec{A}\vec{A}^{-1}\equiv \vec{A}\vec{B}=
\!\left(\!\!\begin{array}{cc}
\vec{P}_{11} & \vec{\rm a}_{12}  \\
\vec{\rm a}^{\top}_{12}  & \vec{\cal A}_{22}  
\end{array}\!\! \right)
\!\left(\!\!\begin{array}{cc}
\vec{H}_{11} & \vec{\rm b}_{12}  \\
\vec{\rm b}^{\top}_{12}  & \vec{\cal B}_{22}  
\end{array}\!\! \right)\!=
\!\left(\!\!\begin{array}{cc}
\vec{I}_{11} & \vec{0}_{12}  \\
\vec{0}^{\top}_{12}  & \vec{\cal I}_{22}  
\end{array}\!\! \right)\equiv \vec{I},
\nonumber \\
&&\!\!\!\!\!\!\!\!\!\!\!\!\!\!\!\!\!\!\!\!\!\!\!\!
\mbox{then: }\;
\vec{\cal B}_{22}=\left(\vec{\cal A}_{22}-
\vec{\rm a}^{\top}_{12}\vec{P}^{-1}_{11} \vec{\rm a}_{12}\right)^{-1}=
\nonumber \\
&&\!\!\!\!\!\!\!\!\!\!\!\!\!\!\!\!\!\!\!\!\!\!\!\!
=\vec{\cal A}^{-1}_{22}+\vec{\cal A}^{-1}_{22}\vec{\rm a}^\top_{12}\left(
\vec{P}_{11}-\vec{\rm a}_{12}\vec{\cal A}^{-1}_{22}\vec{\rm a}^{\top}_{12}\right)^{-1}
\vec{\rm a}_{12}\vec{\cal A}^{-1}_{22},
\nonumber
\end{eqnarray}
whence the rightmost bottom block of the inverse to (\ref{37}) is expressed for 
$i,k,l,j=1\div n-1$ as:
\begin{eqnarray}
&&\!\!\!\!\!\!\!\!\!\!\!\!\!\!\!\!\!\!\!\!\!\!\!\!
\left(\vec{A}^{-1}\right)^{ik}=\left(\vec{\cal B}_{22}\right)^{ik}=
\left(\vec{\cal A}^{-1}\right)^{ik}+
\left(\vec{\cal A}^{-1}\right)^{il} A_{l0}\frac{1}{\alpha}
A_{0j} \left(\vec{\cal A}^{-1}\right)^{jk},
\label{40} \\
&&\!\!\!\!\!\!\!\!\!\!\!\!\!\!\!\!\!\!\!\!\!\!\!\!
\mbox{and since: }\,\left(\vec{\cal A}^{-1}\right)^{ik}=
\frac{(-1)^{i+k}M\left({}^0_0{\;}^i_k\right)}{|{\cal A}_{lj}|}, \quad 
\left(\vec{A}^{-1}\right)^{l0}=
\frac{(-1)^l}{|A_{\mu\nu}|}M\left({}^0_l\right), 
\nonumber 
\end{eqnarray}
the argument of the exponential in (\ref{38}) is also reduced to expression (\ref{36}):
\begin{eqnarray}
&&\!\!\!\!\!\!\!\!\!\!\!\!\!\!\!\!\!\!\!\!\!\!\!\!
\frac{1}{\alpha}B^2_0-\frac{2}{\alpha}B_0B_l
\left(\vec{{\cal A}}^{-1}\right)^{li}\!\!A_{i0}+
\nonumber \\
&&\!\!\!\!\!\!\!\!\!\!\!\!\!\!\!\!\!\!\!\!\!\!\!\!
+B_l\left[\left(\vec{\cal A}^{-1}\right)^{lk}+
\left(\vec{\cal A}^{-1}\right)^{li} A_{i0}\frac{1}{\alpha}
A_{0j} \left(\vec{\cal A}^{-1}\right)^{jk}
\right]B_k =
\nonumber \\
&&\!\!\!\!\!\!\!\!\!\!\!\!\!\!\!\!\!\!\!\!\!\!\!\!
=B_\mu \left(\vec{A}^{-1}\right)^{\mu\nu}B_\nu=\left(B\vec{A}^{-1}B\right).
\nonumber
\end{eqnarray}
A generalization of the integral (\ref{34}) with arbitrary polynomial or smooth 
function similar to the well-known approximations for Euclidian case \cite{Fed} here 
is also straightforward. The imaginary part of $\vec{A}$ (\ref{37}) in integral 
(\ref{34}) appearing for (\ref{23}) as $itg_{\mu\nu}$ preserves convergence of the 
integral and may be included by analytical continuation of the expression (\ref{36}).

\end{document}